# Coherent Power Combining of Four-Way Injection-Locked 5.8-GHz Magnetrons Based on a Five-Port Hybrid Waveguide Combiner

Heping Huang[1,2,3], Bo Yang[2*], *Member IEEE*, Naoki Shinohara[2], Senior *Member, IEEE,* and Changjun Liu[3], Senior *Member, IEEE,*

*Abstract*—A high-efficiency power combining method for four-way 5.8 GHz magnetrons based on the external injection-locking technique is presented in this paper. The method uses a non-isolated, lossless five-port hybrid waveguide combiner for power combining. Meanwhile, the injection locking technology has been applied to magnetrons for achieving coherent power combining. The phase fluctuation of the injection-locked magnetron, without the presence of a phase lock loop, measured nearly ±2.5 degrees. In contrast, when a phase lock loop was introduced, the phase fluctuation reduced significantly to approximately ±0.5 degrees. This phase accuracy can fully meet the requirements of combining experiments. Four magnetrons worked in injection-locked states without phase lock loop. The proposed power combining system is designed, measured, and analyzed. Measurement results show that a high-power combining efficiency of over 95% is achieved by injection-locked magnetron without PLL, with the best efficiency reaching up to 97.7% with phase control of the injected signals. Experimental results reveal that the magnetron phase-pushing effects and the ripple in high-power dc voltage and current have a minor impact of approximately 4% on the combining efficiency.

*Index Terms*— Five-port waveguide combiner, hybrid tee, injection-locking, magnetron, power combining, wireless power transmission (WPT).[1]

## I. INTRODUCTION

POWER combining is emerging in various applications such as wireless communication, wireless power transmission (WPT) [1]-[3], space solar power stations (SPS) [4]-[7], satellite communication [8], and microwave heating [9]. The power capacity of a single CW magnetron (MGT) is mainly limited by its resonant cavity size. Thus, magnetrons at high frequencies, coming with smaller resonant cavities, have lower power capacities compared with low-frequency ones [10],[11]. Therefore, power combining is widely used to achieve high power in high-frequency applications. The basic idea of power combining is simultaneously combining N signals into a single output or vice versa [12]. Over the past decades, there has been increasing research and application of power-combining techniques. Power combining approaches are mainly classified into several categories: 1) chip-level combining [13],[14]; 2) circuit-level combining[15]; 3) spatial combining [16],[17]; 4) hybrid waveguide combining [18]-[23], and 5) combinations of the aforementioned techniques [24]. With the development of large power amplifier chips, several technologies are applied in power combining, such as serial, cascading, impedance optimization, symmetry, and other technologies used at both the chip and circuit levels. Spatial combining is based on active phased arrays. With the development of WPT systems, microwave industrial heating devices, particle accelerators, and so on, there are more and more requirements for large-power microwave generators. Magnetrons are becoming better and more cost-effective choices for those large-power microwave generators. For example, Yang *et al.* achieved an injection-locked 5.8-GHz magnetron active phased array using four independent amplitude and phase-controlled magnetrons [1]. Chen *et al.* developed a 3.5-kW 2.45-GHz microwave- transmitting system based on horn array antennas with four master-slave phase-controlled magnetrons [25]. Hybrid waveguide combining usually provides high power capability and is generally used in vacuum devices.

Magnetrons are widely used in industrial microwave heating applications due to their low cost, high power, and efficiency. However, their unstable phase and frequency negatively affect the precise control of microwave sources in phase array antennas or high-power combining systems

Manuscript received July 28, 2023. This work was supported in part by the Microwave Energy Transmission Laboratory, Research Institute for Sustainable Humanosphere, Kyoto University, and in part by the National Institute of Information and Communications Technology, Japan, under Grant 02401, by the National Natural Science Foundation of China under Grant 62001402 and Grant U22A2015, by the China Scholarship Council under Grant 202008515068, and by the Fundamental Research Funds for the Central Universities, Southwest Minzu University, under Grant ZYN2022028. (*Corresponding author: Bo Yang.*)

Heping Huang is with the Key Laboratory of Electronic and Information Engineering (Southwest Minzu University), State Ethnic Affairs Commission, Chengdu 610225, China, with the Research Institute for Sustainable Humanosphere, Kyoto University, Uji 611-0011, Japan, and also with the School of Electronics and Information Engineering, Sichuan University, Chengdu 610064, China (e-mail: hepinghuang01@outlook.com).

Bo Yang and Naoki Shinohara are with the Research Institute for Sustainable Humanosphere, Kyoto University, Uji 611-0011, Japan (e-mail: yang_bo@rish.kyoto-u.ac.jp; shino@rish.kyoto-u.ac.jp).

Changjun Liu is with the School of Electronics and Information Engineering, Sichuan University, Chengdu 610064, China (e-mail: cjliu@ieee.org).

Color versions of one or more of the figures in this article are available online at http://ieeexplore.ieee.org.

[26],[27]. Injection-locking techniques applied to magnetrons effectively solve the instability problem of magnetrons [28]-[30]. Injection-locked magnetron systems have been investigated in power combining systems and WPT [4]-[7]. Injection-locked magnetron power combining is typically based on 3-dB tees or four-port magic-tees in waveguide combining. Treado *et al.* demonstrated high-power magnetrons driven by an RF-isolated low-power source, achieving a power combining efficiency of 92% using two long-pulse phase-locking magnetrons combined with a 3-dB hybrid coupler [31]. Liu *et al.* successfully combined two-way and four-way injection-locked 2.45-GHz magnetrons based on a 3-dB waveguide divider, achieving a combining efficiency of over 90% [10],[21]. Park *et al.* performed power combining experiments using two 2.45-GHz identical magnetrons and achieved an efficiency of approximately 93% [32].

In this work, a stable and highly efficient four-way injection-locked 5.8-GHz power combining experiment was conducted. We proposed a high-efficiency power combining method consisting of four-way injection-locked magnetron systems, and the power combining experiment was based on a compact five-port hybrid combiner. It achieved high combining efficiencies for both low- and high-power scenarios. Only the injected signals were used for low-power combining, and the combining efficiency for the four-way injected signals without injection to the magnetron was above 96%. Building upon the well-controlled technology of injection-locked magnetrons studied in our previous work [1], in this works we study the phase characteristics of the phase locked magnetron by improving the reducing the dc high voltage power supply ripple, external signal injection locked technology, or closed-loop phase locking and so on. Our focus revolves around enhancing various aspects, such as mitigating the ripple in the dc high voltage power supply, implementing external signal injection locked technology, and employing closed-loop phase locking, among other strategies. Upon activation of the four-way magnetrons with the combining of injected signals, the observed efficiency in combining magnetron output power surpassed an impressive 93%. By adjusting the phase shifter, the high-power combining efficiency was over 95%, with the best efficiency reaching 97.7%.

## II. COMBINING EFFICIENCY CHARACTERISTICS

Consider a power combiner that works well with an n-way linear independent signal source network, where only one output of the combiner is the final combined power output. The combining efficiency $\eta_{com}$ of the n-way signals is defined as the ratio between the output power $P_{com}$ of the combiner and the arithmetic sum of the powers $P_{av}$ of the n individual signal sources [33]-[36] and given by

$$\eta_{com} = \frac{P_{com}}{\sum_{i=1}^{n} P_{av,i}} \times 100\% . \quad (1)$$

For a selected combiner, the maximum efficiency $\eta_{max}$ is its intrinsic merit [33],[34]. The maximum efficiency is guaranteed when the input signal amplitudes and phases are identical, aligning with the intrinsic property of the combiner [33],[34]. However, amplitude and phase differences will exist between the ports when the power and phase of each port's signal are not identical. In this case, if the power and phase available from each of the individual signal sources being combined connected to the *i*th port of the combiner are denoted by $P_i$ and $\theta_i$, respectively, the combining output power $P_{com}$ is the vectorial sum of the arithmetic sums of the powers [34]:

$$P_{com} = \eta_{max} \frac{1}{n} \left[ \left( \sum_{i=1}^{n} \sqrt{P_i} \cos\theta_i \right)^2 + \left( \sum_{i=1}^{n} \sqrt{P_i} \sin\theta_i \right)^2 \right]. \quad (2)$$

Then, the combining efficiency from (1) can be determined as follows [34]:

$$\eta_{com} = \eta_{max} \frac{\left[ \left( \sum_{i=1}^{n} \sqrt{P_i} \cos\theta_i \right)^2 + \left( \sum_{i=1}^{n} \sqrt{P_i} \sin\theta_i \right)^2 \right]}{n \sum_{i=1}^{n} P_{av,i}} \times 100\% . \quad (3)$$

$$\leq 1 \ (always)$$

The combining efficiency reaches the maximum value affected by two main factors [33]. One is the characteristics of the combiner itself, including the matching and isolation, symmetry, power dissipation, bandwidth, and power capacity. Another factor is the characteristics of the signals to be combined, including the flexibility in adjusting the amplitudes and phases of the signals. Additionally, in practical applications, it is necessary to reduce the power losses occurring in the combining.

For a power combining system based on injection-locked magnetrons, the selection of the combiner is crucial. In selecting the appropriate power-combining component, three primary factors are carefully weighed.

1) Power capacity: when dealing with high-power vacuum devices, the predominant choice is waveguide. In particular, magnetrons, which possess substantial power, make waveguide devices the ideal selection.

2) Symmetry and low power dissipation: maximizing efficiency hinges on analyzing combiner losses and reflections. Therefore, the chosen combiner must exhibit symmetry and uniform power distribution, which leads to possible maximum power combining efficiency. Additionally, using low-loss devices is highly advisable to minimize power dissipation.

3) Cascade length: power combining efficiency experiences a significant decline since power loss increases exponentially with transmission line length. Consequently, this imposes a practical constraint on the viability of serial and corporate power combiners, allowing them to be employed efficiently only within a limited number of stages [12].

## III. Quasi-Symmetric Hybrid Combiner

Within waveguide combiners, two-way combining is commonly achieved using E-plane tee, H-plane tee, or magic-tee configurations [9], while four-way combining requires multiple tees to cascade [21]. A multiple-ports compact combiner reduces the cascade levels, volume, and weight of the whole system. A five-port combiner was applied to achieve four-way magnetron output power in the experiment. The design concept of the quasi-symmetric combiner was derived from a two-way combiner consisting of a 3-dB E-plane tee and a 3-dB H-plane tee. Fig. 1 shows the quasi-symmetric structure of the combiner. A rectangular waveguide with ports 2-5 was assembled with a 45° corner. It should be noted that all diagonal elements of the scattering matrix (S-matrix) of an E- or H-plane T-junction cannot be simultaneously zero as the tee junction cannot be ideally matched to all other arms simultaneously [37],[38]. The S-matrix of an E-plane tee can be derived considering port 3 as matched. Ports 2 and 3 serve as the input ports of the H-T combiner, while ports 4 and 5 act as the input ports of the E-T combiner. Port 1 is the output port.

Port 1 is the power combining output port, and the electronic field configuration of the combiner is depicted in Fig. 2. To achieve a quasi-symmetric 3-dB H-plane tee, ports 2 and 3 were combined with port 1 [38]. Ports 4 and 5 were assembled with port 1 to achieve the 3-dB E-plane tee [38]. The S-matrix of ports 1, 2, and 3, and the S-matrix of ports 1, 4, and 5 can be respectively represented as follows:

$$S_{E\text{-}T}=\begin{bmatrix}0 & -\frac{1}{\sqrt{2}} & \frac{1}{\sqrt{2}} \\ -\frac{1}{\sqrt{2}} & \frac{1}{2} & \frac{1}{2} \\ \frac{1}{\sqrt{2}} & \frac{1}{2} & \frac{1}{2}\end{bmatrix},\ S_{H\text{-}T}=\begin{bmatrix}0 & \frac{1}{\sqrt{2}} & \frac{1}{\sqrt{2}} \\ \frac{1}{\sqrt{2}} & \frac{1}{2} & -\frac{1}{2} \\ \frac{1}{\sqrt{2}} & -\frac{1}{2} & \frac{1}{2}\end{bmatrix}. \quad (4)$$

If we assume that the E- and H-plane T-junction of the combiner is symmetric and lossless, respectively, and port 1 has perfect impedance matching, the S-matrix of the five-port combiner [39] is

$$\mathbf{S}=\begin{bmatrix}\mathbf{0} & \mathbf{S}_{12} & \mathbf{S}_{12} & \mathbf{S}_{14} & \mathbf{S}_{12} \\ \mathbf{S}_{12} & \mathbf{S}_{22} & \mathbf{S}_{23} & \mathbf{S}_{24} & \mathbf{S}_{25} \\ \mathbf{S}_{12} & \mathbf{S}_{23} & \mathbf{S}_{22} & \mathbf{S}_{34} & \mathbf{S}_{35} \\ \mathbf{S}_{14} & \mathbf{S}_{24} & \mathbf{S}_{34} & \mathbf{S}_{44} & \mathbf{S}_{45} \\ \mathbf{S}_{12} & \mathbf{S}_{25} & \mathbf{S}_{35} & \mathbf{S}_{45} & \mathbf{S}_{44}\end{bmatrix}. \quad (5)$$

The maximum efficiency of the combiner is an intrinsic property. To ensure the maximum efficiency of the combiner, it is essential to match the amplitude and phase characteristics of the signals to be combined. The five-port power combiner's scattering parameters (S-parameters) were measured and simulated, as shown in Fig. 3.

The amplitude and phase differences of the input port and the output power of the five-port combiner were specially designed. To combine the output of the same four sources and to achieve the maximum power combining efficiency, it is essential to ensure that the phase and amplitude characteristics between each port are suitable. The coupling of the waveguide combiner can be challenging due to the junctions of the waveguide tees, which are often poorly matched devices. Unequal power distribution can impact the coupling between each input port and subsequently affect the combining efficiency. Therefore, the intrinsic amplitude and phase relationships between port 1 and the other ports must satisfy the following conditions:

$$|S_{12}|=|S_{13}|=|S_{15}|=|S_{14}|=1/2. \quad (6)$$

$$\varphi_{12}=\varphi_{13}=\varphi_{15},\ \varphi_{14}=180^{\circ}-\varphi_{15}. \quad (7)$$

The measured and simulated S-parameters are depicted in Fig. 4 and Fig. 5. When port 1 of the power combiner is well matched, both the simulated and experimental return losses ($|S_{11}|$) are better than −20 dB. However, the amplitude satisfies (6) near 5.8 GHz. The measured phase characteristics of $S_{21}$, $S_{31}$, $S_{41}$, and $S_{51}$ are shown in Fig. 5. The phase characteristics fully satisfied (6) and (7).

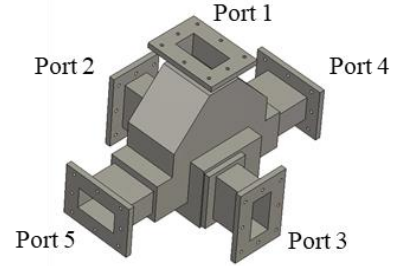

Fig. 1. Model of the four-way combiner.

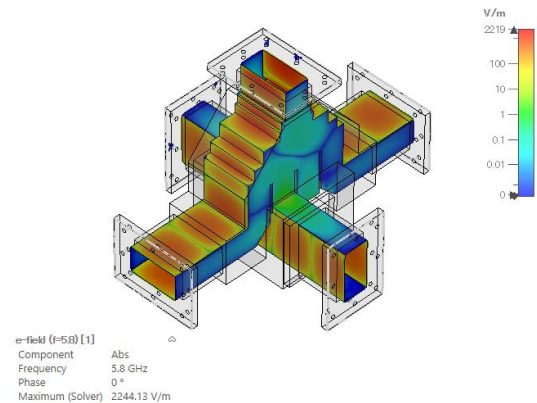

Fig. 2. Electronic field of the four-way combiner.

The flexibility in adjusting the amplitudes and phases of the signals to be combined is determined by the well-controlled injection-locked magnetrons, which were studied in our previous work [1]. The amplitude and phase alignment between the magnetron output ports and the combiner input port is essential for achieving the best power combining efficiency. The microwave's phase $\varphi$ and amplitude $V_i$ at each port of the power combiner are shown in Fig. 6. The microwave at port 4 is out of phase with ports 2, 3, 4, and 5 since port 4 is in the H-plane, unlike the other input ports in the E-plane. Fig. 6 (a) illustrates the ideal state for maximum power combing efficiency. The microwave at each input port 2, 3, 4, and 4 has the same phase and amplitude, except port 4 is out of phase. Fig. 6 (b) shows a practical state. The microwave at each input port has various amplitudes, and each input port's phase is not identical. Then, the power combining efficiency decreases.

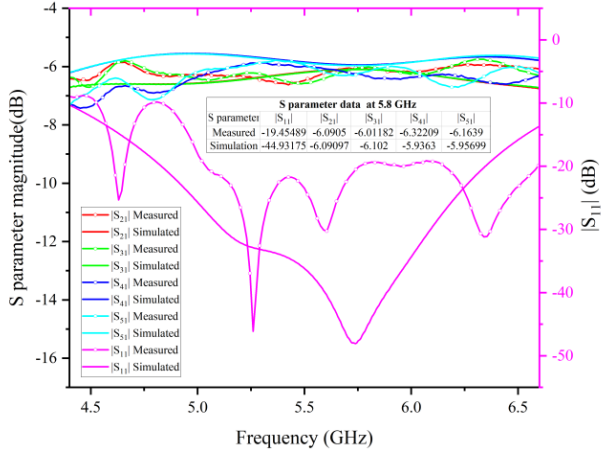

Fig. 3. Simulated and measured *S*-parameters between port 1 and ports 2−5.

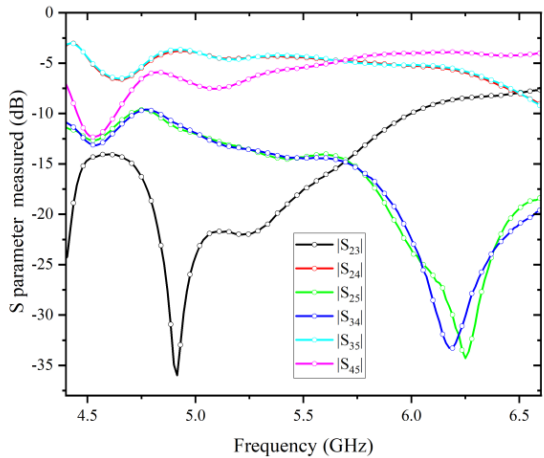

Fig. 4. Measured *S*-parameter characteristics among ports 2−5.

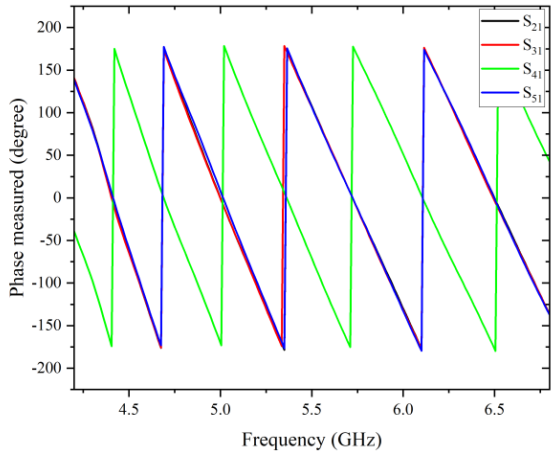

Fig. 5. Measured phase characteristics between port 1 and ports 2−5.

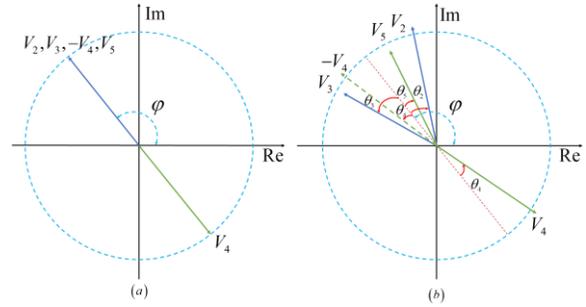

Fig. 6. Phase and amplitude of microwave at each input port. (a) Ideal power combing state. (b) Practical power combining state.

The S-parameter of the combiner determines the phase relationship. The power and phase of the input signals are controlled by the individual sources, and the coupling of the waveguide combiner, as well as the matching between the individual sources and the combiner, affects the actual input power into the combiner. If all the matching and connection problems of the individual sources and the combiner have fulfilled the requirements, the combining efficiency is equal to the maximum efficiency of the combiner. However, it is challenging to maintain identical power and identical phases in the case of multiple magnetrons. Therefore, a specific difference in power or phase is acceptable and can still yield satisfactory results. In practical applications, the amplitude and phase characteristics of the signals are another factor in combining efficiency. Instead, there are three other common scenarios: 1) unequal power and unequal phases, 2) identical power and unequal phases, and 3) unequal power and identical phase [34]. For an *n*-way power combining system, with the available reference power and phase with standard power deviation $\Delta G$ (in dB) and phase deviation $\Phi$ (in degrees), the efficiency degradation should be simplified as a function[34],[40]:

$$\eta_{com} = \frac{4 \times 10^{\Delta G/10} \cos^2 \Phi}{(1+10^{\Delta G/10})^2}. \quad (8)$$

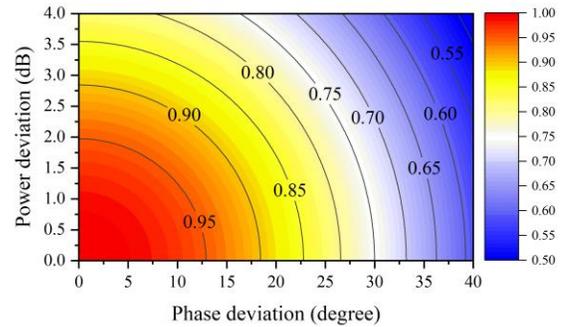

Fig. 7. Power combining efficiency with power and phase deviations.

Fig. 7 shows the contours of the minimum combining efficiency of the power and phase deviations. The larger the standard power and phase deviations, the lower the power combining efficiency. If the power combining efficiency is greater than 95%, the phase variation is lower than 12°, and the power variation allows some tradeoffs up to approximately 3 dB. Therefore, the phase variation is essential in the combining experiment.

## IV. PHASE-LOCKED MAGNETRON INVESTIGATION

A 5.8 GHz phase-locked magnetron system was meticulously designed to facilitate comprehensive investigations into phase-locking dynamics, as illustrated in Fig. 8. To optimize phase stability in the phase-locked magnetron, various strategies were meticulously considered. These included the reduction of dc high voltage power supply ripple, the integration of external signal injection-locked technology, and the implementation of closed-loop phase locking, among other methodologies [1]. The experimental setup involved the injection of an external signal, generated by a signal generator and manipulated through a phase shifter and power amplifier, ultimately directed into the magnetron via a circulator. The accompanying diagram delineates the system's conditions both before (depicted by the blue line, labeled 'a') and after the phase-locked loop (PLL) circuit (depicted by the green line, labeled 'b') was engaged together. The assessment of the system's performance involved measuring the frequency spectrum and power output, accomplished using a spectrum analyzer (Agilent N9020A) and a power meter (Agilent N1914A).

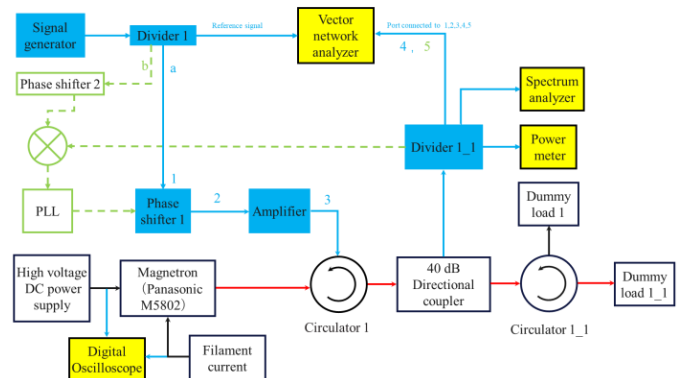

Fig. 8. Phase measurement of the phase-locked magnetron.

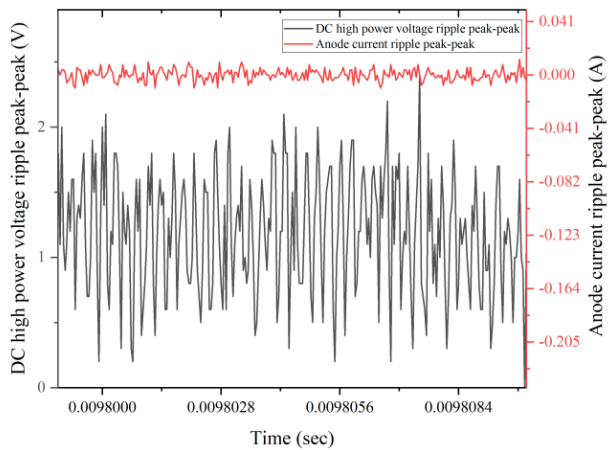

Fig. 9. Phase measurement of the magnetron's dc power supply.

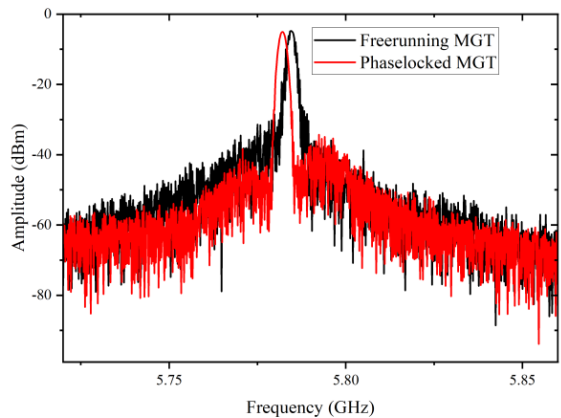

Fig. 10. Spectrum measurement of magnetron before and after phase-locked.

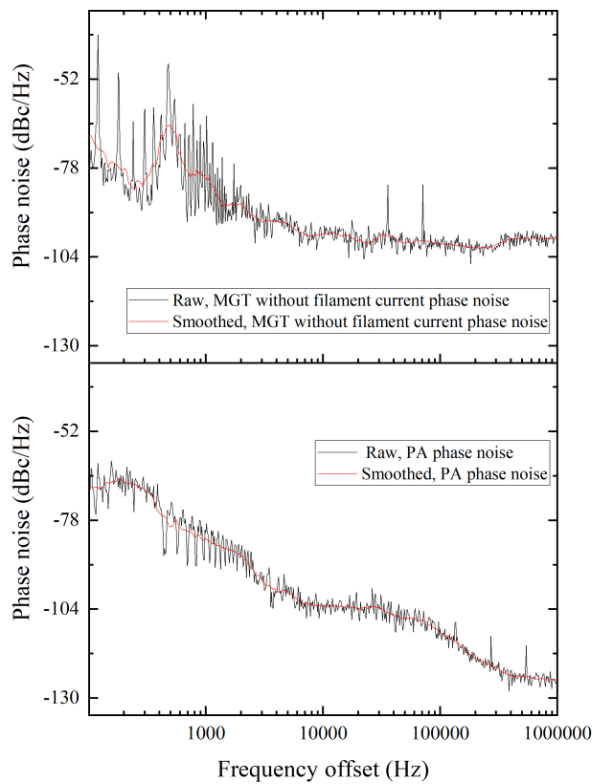

Fig. 11. Phase noise of the phase-locked magnetron.

The ripple of the dc high-power supply and anode current was subjected to measurement. The magnetron system operated with a high dc voltage of 4480 V, while the specified filament voltage and current were specified as 3.3 V and 7.4 A by ac 3.35 V. Notably, the peak-to-peak fluctuation of the dc high-power supply and anode current ripple was found to be below 2.9 V and 0.04 A, as illustrated in Fig. 9. Fig. 10 presents the frequency spectra of the magnetron under free-running and phase-locked conditions. Additionally, Fig. 11 displays the phase noise characteristics of both the injected signal and the phase-locked magnetron output. These results provide a comprehensive understanding of the system's performance and stability.

A reference signal sourced from the signal generator (SG) is divided by divider 1 and utilized in phase difference measurements conducted by a vector network analyzer (Agilent N5242A). The phase difference between the reference signal and signals at positions 1-5 were measured. As depicted in Fig. 12, the phase fluctuation between the signal generator output via the line connected with a phase shifter (position 2 in Fig. 8) and the reference signal was approximately ±0.2 degrees as shown with green line in Fig. 12. The phase fluctuation was approximately ±0.4 degrees (as shown with blue line in Fig. 12) between the signal from divider 1 amplified through a power amplifier connected to the injecting port (position 3 in Fig. 8), The phase fluctuation was approximately ±2.5 degrees as shown with red line in Fig. 12 between the injection-locked magnetron (position 4 in Fig. 8) and the reference signal. It is worth noting that these observed phase variations meet the specified requirements, particularly if the power combining efficiency exceeds 95%. This data supports the conclusion that the phase stability achieved is sufficient for phase variation criteria of the combining experiment.

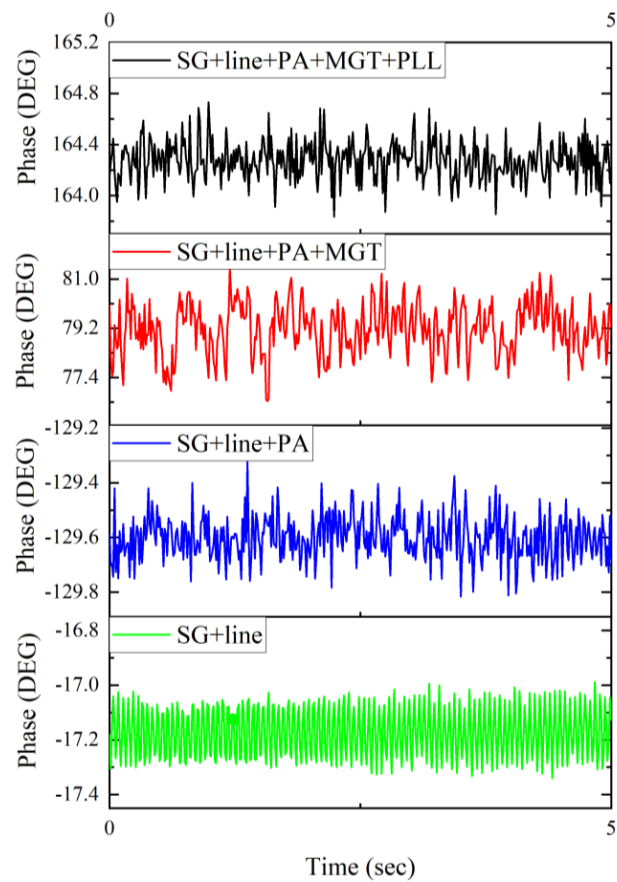

Fig. 12. Phase measurement of the phase-locked magnetron.

For applications demanding superior phase characteristic, the introduction of a phase-locked loop (PLL) is recommended to achieve a closed-loop phase-locked magnetron system, Prior to the operation of the phase-locked loop circuit, an external signal from the signal generator, manipulated through a phase shifter and amplifier, was injected into the magnetron via a circulator, denoted by blue blocks and route a in Fig. 8. A closed-loop with a phase-locked loop was implemented, following route b, marked with green blocks in Fig. 8. The magnetron's output was fed back into the phase-locked loop using a mixer before entering the phase shifter. Subsequently, the phase difference between magnetron output (position 5 in Fig. 8) and the reference signal was measured, and the results is nearly ±0.5 degree, as shown in Fig. 12 with black line. In subsequent combination experiments, an injection-locked magnetron without a phase-locked loop was employed to streamline system complexity and achieve high combining efficiency.

## V. EXPERIMENT AND DISCUSSION OF RESULTS
### A. Experiment system

The block diagram of the power combining system, as shown in Fig. 13, illustrates the system's configuration. An external signal is divided by a four-way power divider, and each division is connected to a respective phase shifter (designated as φ). Each divided signal is then amplified by a power amplifier (R&K CA5800BW50-4040R RF power amplifier) and directly injected into each magnetron through a circular waveguide. The injected signal has a lower power level compared to the magnetron output, making phase control easier. The phase shifter adjusts the phase of the injected signal to align with the magnetron output signal. The magnetrons are numbered from 2−5, corresponding to the number of the combiner ports. The hardware of the injection-locked magnetron subsystem is depicted in Fig. 14.

In the experiment, four commercial magnetrons (Panasonic M5802) were used. The high dc voltage of the magnetrons was 4480 V, and the filament voltage and current were specified as 3.3 V and 7.4 A, respectively. Each magnetron was powered by a high dc voltage and a filament current. The microwave output of each magnetron was detected by a power meter (Agilent E4419B and Agilent N1914A) through a circular directional coupler at each side. The detected signals were then connected to the five-port waveguide combiner.

The four-way magnetron outputs were combined using the waveguide combiner. The combined output was measured using a power meter (HP EPM-442A) and a spectrum analyzer (Agilent E4440A) to obtain the combined power and spectrum. Based on the above schematic, the four-way 5.8-GHz continuous wave magnetron microwave power combining system was constructed, and a photograph of the system is shown in Fig. 15.

### B. External Signal Power Combining

Table I presents the experimental results of the combined injected signals under the identical in-phase condition. The signals input to ports 2 to 5 are from the injection-locked subsystem with all the magnetrons turned off. The maximum efficiency is achieved and recorded by adjusting the phase shifter of each injected signal in case 1. From case 2 to 6 with power level increased step by step but keep the phase shifter setting in Case 1, the maximum combining efficiency is recorded, respectively. As the power of the injected signals is increased through adjustment of the signal generator, the combining efficiency slightly fluctuates and remains above 96%. It should be noted that the phase between the input signals and the combiner's inherent characteristics needs to be coordinated during the measurement process.

Table II presents experimental results obtained using various power levels of injected microwave with the magnetrons turned off as well. We tuned the phase shifters in each case to receive and record the maximum power combining efficiency. As the power of the injected signal increases, the phase shifters are adjusted to obtain the maximum output power at port 1 of the combiner, thereby achieving the best efficiency in each case. The overall efficiency is above 97%. The experimental system can maintain a stable and high combining efficiency by comparing the two sets of data and the measurement processes.

The rated power of the power amplifier was 10 W. To ensure the injection-locked state while increasing the magnetron output, some of the injected signals are near the maximum value of the power amplifier. For example, in case 6, the data show that the power of amplifiers 2 and 4 has reached saturation.

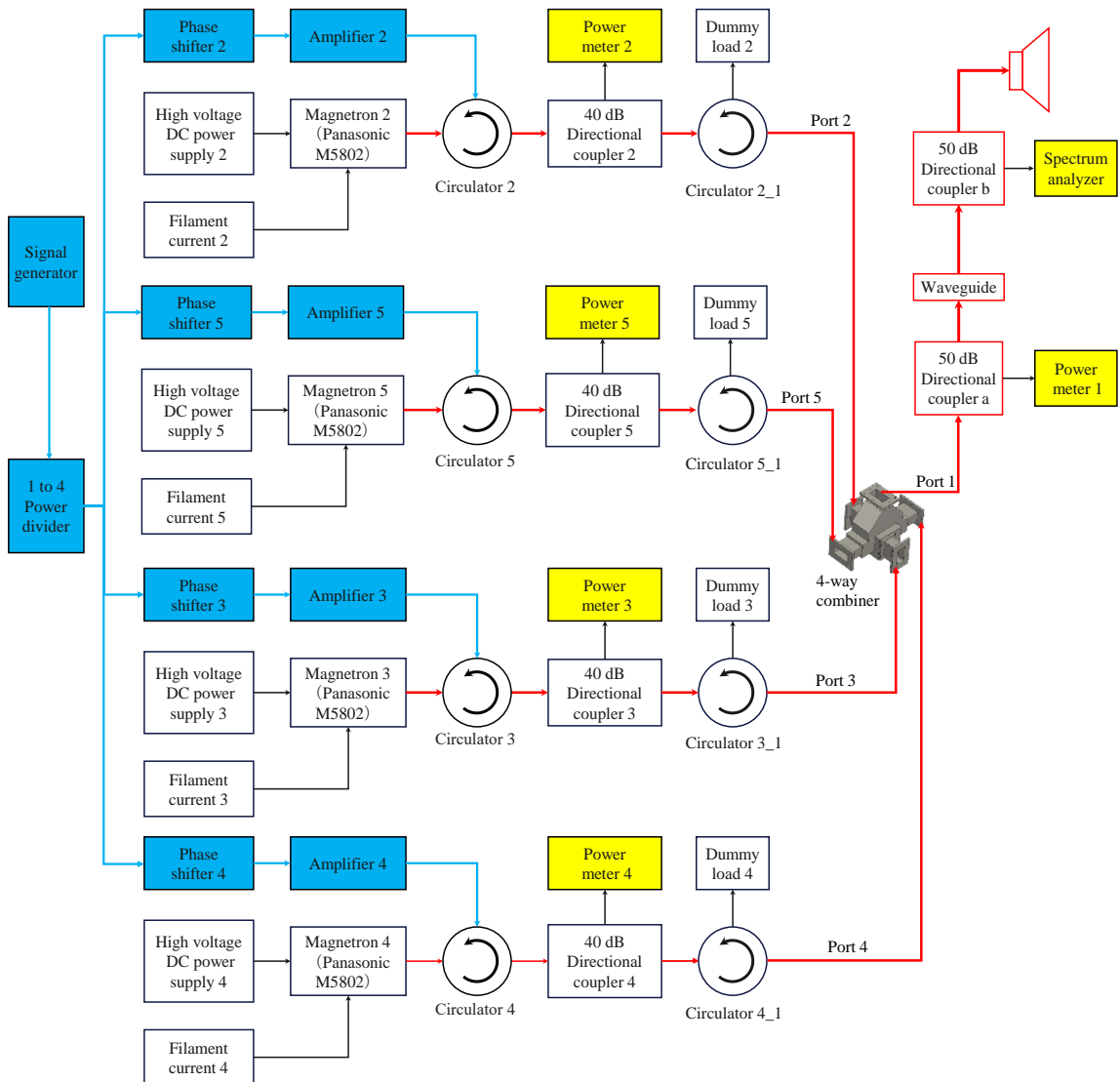

Fig. 13. Diagram of the four-way injection-locked magnetron power combining system.

## C. Magnetron Output Power Combining

When the system maintains the combined state of the injected signals at maximum efficiency, the magnetrons are turned on one by one. Table III presents the power combining of the magnetron outputs under the same phase condition. Table III shows experimental results obtained using various power levels of injected microwave under identical phase conditions. We tuned the phase shifters in Case 1 to receive and record the maximum power combining efficiency. Maximum power combining efficiency of each case has been recorded. The filament currents must be adjusted to zero to reduce the noise of the magnetron output after the four magnetrons work normally, which is more conducive to the subsequent injection locking process. The injection locking status works well, as indicated by the locking spectra of the four-way magnetron output shown in Fig. 16. Case 2-6 in Table III, the phase shifter was not adjusted as the magnetron output power increased. The power of the magnetrons was increased by adjusting the dc high-power anode voltage and current.

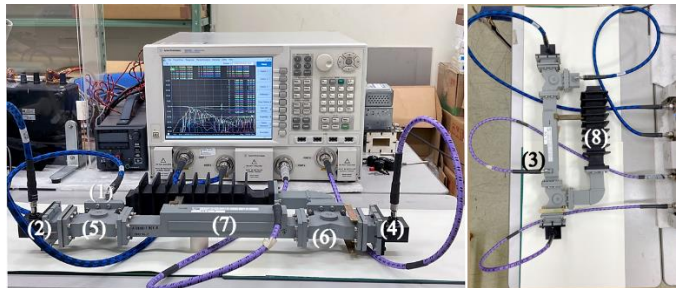

Fig. 14. Parts of the injection-locked magnetron subsystem: (1) Port for the injected signal; (2) port connected to the magnetron; (3) port for connecting the power meter; (4) port connected to the combiner; (5) and (6) circulars; (7) directional coupler; and (8) dummy load.

The four combined magnetrons in this technology used a non-isolated waveguide combiner, which reduces the energy loss within the combiner. Simultaneously, each magnetron maintains a stable injection-locked state with a synchronized and stable phase, resulting in a high combining efficiency. As shown in Table III, the power combining efficiency is above 93%, which is slightly lower than the results shown in Table I, indicating that the impact of the injection locking phase shift on the combining efficiency is approximately 4%. Tables III and IV demonstrate that the process of injection locking the magnetrons affects the output power. The injection-locked magnetrons exhibited a phase-pushing effect, as observed in the results, and it is also influenced by the ripples in dc high-power anode voltage and current.

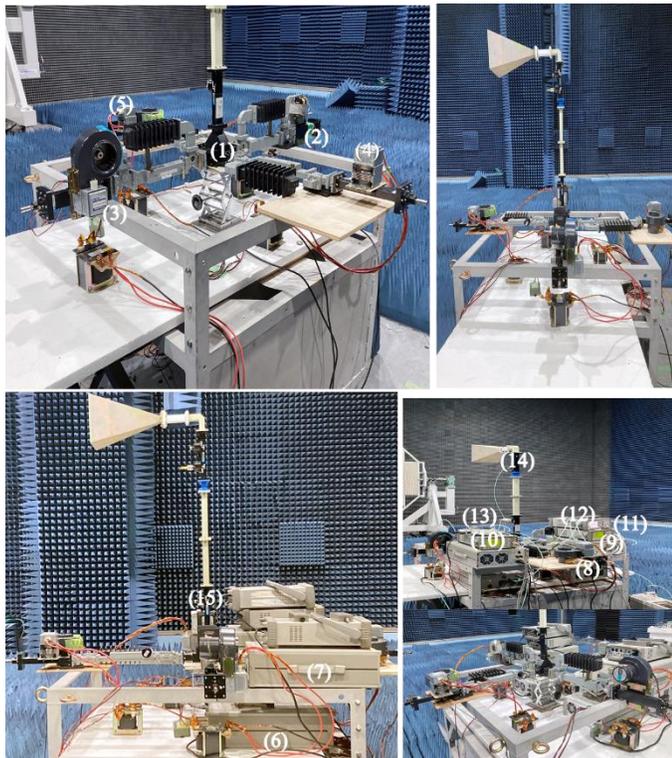

Fig. 15. Photographs of the combining experiment: (1) Five-port combiner; (2)–(5) parts of injection-locked magnetron subsystem; (6)–(9) power amplifiers, (10)–(12) power meters; (13) phase shifter; (14) 50-dB directional coupler b; and (15) 50-dB directional coupler a.

TABLE I
POWER COMBINING OF THE INJECTED SIGNALS UNDER THE IDENTICAL PHASE CONDITION

| Case#<br>Output (W) | 1 | 2 | 3 | 4 | 5 | 6 |
|---|---|---|---|---|---|---|
| Port 2 | 1.26 | 3.16 | 3.43 | 6.01 | 7.98 | 10.06 |
| Port 3 | 1.27 | 3.20 | 3.33 | 5.64 | 7.62 | 9.35 |
| Port 4 | 1.31 | 3.23 | 3.41 | 5.63 | 7.62 | 10.61 |
| Port 5 | 1.10 | 2.91 | 3.12 | 4.97 | 6.57 | 7.13 |
| Port 1 | 4.80 | 12.30 | 12.90 | 21.40 | 28.80 | 35.90 |
| $\eta_{com}$ (%) | 97.1 | 98.4 | 97.0 | 96.1 | 96.6 | 96.6 |

TABLE II
POWER COMBINING OF THE INJECTED SIGNALS UNDER THE BEST PHASE CONDITION

| Case#<br>Output (W) | 1 | 2 | 3 | 4 | 5 | 6 |
|---|---|---|---|---|---|---|
| Port 2 | 1.24 | 2.47 | 3.86 | 5.02 | 7.75 | 9.91 |
| Port 3 | 1.55 | 3.01 | 4.60 | 5.89 | 8.80 | 8.88 |
| Port 4 | 1.34 | 2.62 | 3.98 | 5.04 | 7.72 | 8.42 |
| Port 5 | 1.14 | 2.34 | 3.71 | 4.81 | 6.26 | 6.41 |
| Port 1 | 5.16 | 10.20 | 15.80 | 20.30 | 29.80 | 33.00 |
| $\eta_{com}$ (%) | 97.9 | 97.7 | 97.8 | 97.7 | 97.6 | 98.1 |

TABLE III
THE POWER COMBING OF THE MAGNETRON OUTPUTS UNDER IDENTICAL PHASE CONDITION

| Case #<br>Output (W) | Case 1 | Case 2 | Case 3 | Case 4 | Case 5 | Case 6 |
|---|---|---|---|---|---|---|
| current (mA) | 100 | 110 | 120 | 130 | 140 | 150 |
| Port 2 | 227 | 253 | 279 | 304 | 331 | 357 |
| Port 3 | 217 | 240 | 266 | 293 | 317 | 343 |
| Port 4 | 199 | 220 | 239 | 258 | 275 | 296 |
| Port 5 | 199 | 219 | 240 | 262 | 282 | 305 |
| Port 1 | 791 | 894 | 984 | 1060 | 1141 | 1220 |
| $\eta_{com}$ (%) | 93.9 | 95.9 | 96.0 | 94.8 | 94.6 | 93.7 |

The high dc power supply is at 4480 V at each case.

TABLE IV
THE POWER COMBINING OF MAGNETRON OUTPUTS UNDER THE BEST PHASE CONDITION

| Case#<br>output (W) | Case 1 | Case 2 | Case 3 | Case 4 | Case 5 | Case 6 |
|---|---|---|---|---|---|---|
| Port 2 | 230 | 251 | 274 | 297 | 318 | 331 |
| Port 3 | 244 | 268 | 294 | 317 | 343 | 360 |
| Port 4 | 212 | 233 | 256 | 279 | 300 | 317 |
| Port 5 | 179 | 196 | 216 | 234 | 252 | 263 |
| Port 1 | 827 | 916 | 1010 | 1087 | 1172 | 1242 |
| $\eta_{com}$ (%) | 95.6 | 96.6 | 97.1 | 96.4 | 96.6 | 97.7 |

The high dc power supplies and currents are different at each case.

In Table IV, the combining state of the injected signals is maintained, and we measured the power combining of the magnetrons under the respective best phase conditions. In each case, all four magnetrons achieved injection locking. It is important to note that one magnetron may lose an injection-locked state at lower output power levels even with a large injected signal. However, by increasing the output power of the magnetron or reducing the oscillation noise, the magnetron can be brought into a stable injection-locked state. The phase shifters were adjusted accordingly after increasing the magnetron output power each time. The results are shown in Table IV, which demonstrates the maximum combining efficiency of the four-way magnetrons achieved in the experiment. The combining efficiency is over 95%, and the best efficiency is up to 97.7%.

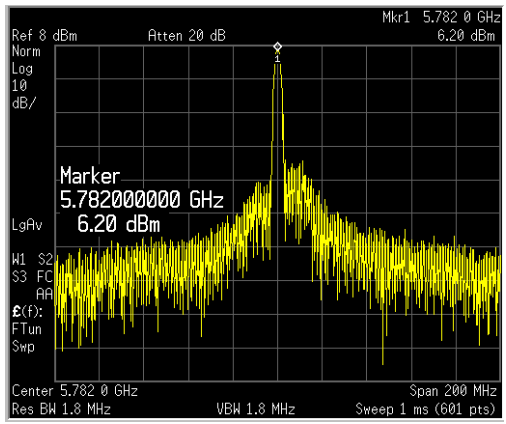

Fig. 16. Spectrum of the four-way magnetron power combining signal.

## VI. CONCLUSION

In the four-way injection-locked 5.8-GHz power combining system, a compact, non-isolated, low-loss hybrid waveguide combiner with five ports was used, resulting in a high combining efficiency. The phase characteristics of the injection-locked magnetron were thoroughly investigated. The phase fluctuations observed in the injection-locked magnetron were approximately ±2.5 degrees in the absence of a phase lock loop and were significantly reduced to ±0.5 degrees with the incorporation of a phase lock loop. This high efficiency was achieved through the combination of injection-locked magnetrons. When the signals were injected, and the magnetrons were in the injection-locked state, coherent power combining was successfully accomplished with and without phase adjustment. Adjusting the phase shifter achieved a high-power combining efficiency of over 95%, with the best efficiency reaching up to 97.7%. The magnetron phase pushing effect, as well as the ripple in dc high-power voltage and current, has an impact of approximately 4% on the power combining efficiency.

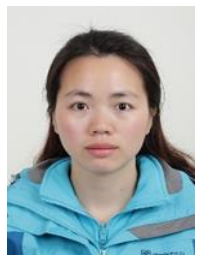

**Heping Huang** received the B.S. degree in communication engineering and the M.S. degree in circuit and systems from Hunan Normal University, Changsha, China, in 2010 and 2013, respectively. She is received the Ph.D. degree in radio physics from Sichuan University, Chengdu, China. Since 2017, she has been a faculty member with Southwest Minzu University, Chengdu, China. Her current research interests include microwave heating and its industrial applications, microwave circuits, and techniques.

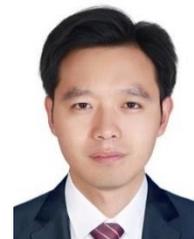

**Bo Yang** (S'16–M'20) received the BE degree in electronic information engineering from China University of Petroleum, Qingdao, China in 2008 and the ME and Ph.D. degrees in electrical engineering from Kyoto University, Japan, in 2018 and 2020. From 2019 to 2021, he conducted research on high-power wireless power transfer systems, supported by Research Fellowships for Young Scientists sponsored by the Japan Society for the Promotion of Science (JSPS). He is currently researching high-power microwave wireless power transmission at the Research Institute for Sustainable Humanosphere, Kyoto University. He is the Associate Editor of the Space solar power and wireless transmission. From 2008 to 2015, he was an RF engineer with the DAIHEN Group Qingdao.

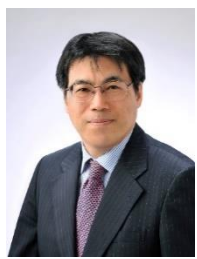

**Naoki Shinohara** (M'06–SM'19) received the B.E. degree in electronic engineering the M.E. and Ph.D. (Eng.) degrees in electrical engineering from Kyoto University, Japan, in 1991, 1993, and 1996, respectively. He was a research associate in the Radio Atmospheric Science Center, Kyoto University from 1996, and there he was an associate professor since 2001. He was an associate professor in Research Institute for Sustainable Humanosphere, Kyoto University, by recognizing the Radio Science Center for Space and Atmosphere in 2004. From 2010, he has been a professor in Research Institute for Sustainable Humanosphere, Kyoto University. He has been engaged in research on Solar Power Station/Satellite and Microwave Power Transmission system. He is IEEE Distinguish Microwave lecturer, IEEE MTT-S Technical Committee 26 (Wireless Power Transfer and Conversion) vice chair, IEEE MTT-S Kansai Chapter TPC member, IEEE Wireless Power Transfer Conference advisory committee member, international journal of Wireless Power Transfer (Cambridge Press) executive editor, Radio Science for URSI Japanese committee C member, past technical committee chair on IEICE Wireless Power Transfer, Japan Society of Electromagnetic Wave Energy Applications vice chair, Wireless Power Transfer Consortium for Practical Applications (WiPoT) chair, and Wireless Power Management Consortium (WPMc) chair.

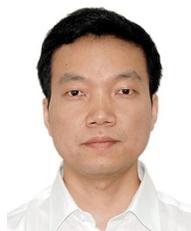

**Changjun Liu** (Senior Member, IEEE) received the B.S. degree from Hebei University, China, in 1994, and the M.S. and Ph.D. degree from Sichuan University, China, in 1997 and 2000, respectively. From 2000 to 2001, he was a Post-Doctoral Researcher with Seoul National University, Korea. From 2006 to 2007, he was a Visiting Scholar with Ulm University, Germany. Since 1997, he has been with the Department of Radio Electronics, Sichuan University, where he has been a professor since 2004. He has authored one book and more than 100 articles. His current research interests include microwave power combining, microwave wireless power transmission, and microwave power industrial applications.

Dr. Liu was a recipient of several honors, such as the outstanding reviewer for the IEEE MTT-S from 2006 to 2010, support from the MOE under the Program for New Century Excellent Talents in University, China from 2012 to 2014; the Sichuan Province Outstanding Youth Fund, from 2009 to 2012; and named by Sichuan Province as an Expert with Outstanding Contribution.